\documentclass[aps,pra,twocolumn,superscriptaddress,showkeys]{revtex4-1}

\usepackage{graphicx}
\usepackage{amssymb}
\usepackage{subfigure}
\usepackage{amsmath}
\usepackage{amsfonts}
\usepackage{bm}
\usepackage{color}
\usepackage[dvipsnames]{xcolor}
\usepackage{physics}
\usepackage{gensymb}

\newcommand{\e}{\textup{e}}

\begin{document}

\title{Full Thermalization of a Photonic Qubit}

\author{A. G. de Oliveira}
\affiliation{Departamento de F\'{i}sica, Universidade Federal de Santa Catarina, CEP 88040-900, Florian\'{o}polis, SC, Brazil}
\author{R. M. Gomes}
\affiliation{Institute of Physics, Federal University of Goi\'{a}s, 74690-900, Goi\^{a}nia, GO, Brazil}
\author{V. C. C. Brasil}
\affiliation{Departamento de F\'{i}sica, Universidade Federal de Santa Catarina, CEP 88040-900, Florian\'{o}polis, SC, Brazil}
\author{N. Rubiano da Silva}
\affiliation{Departamento de F\'{i}sica, Universidade Federal de Santa Catarina, CEP 88040-900, Florian\'{o}polis, SC, Brazil}
\author{L. C. C\'{e}leri}
\affiliation{Institute of Physics, Federal University of Goi\'{a}s, 74690-900, Goi\^{a}nia, GO, Brazil}
\affiliation{Department of Physical Chemistry, University of the Basque Country UPV/EHU, Apartado 644, E-48080 Bilbao, Spain}
\author{P. H. Souto Ribeiro}
\email{p.h.s.ribeiro@ufsc.br}
\affiliation{Departamento de F\'{i}sica, Universidade Federal de Santa Catarina, CEP 88040-900, Florian\'{o}polis, SC, Brazil}

\begin{abstract}

The generalized amplitude damping (GAD) quantum channel implements the interaction between a qubit and an environment with arbitrary temperature and arbitrary interaction time. Here, we implement a photonic version of the GAD for the case of infinite interaction time (full thermalization). We also show that this quantum channel works as a thermal bath with controlled temperature.

\end{abstract}


\maketitle


\section{Introduction}

Photon polarization is a very useful degree of freedom for the realization of several tasks in Quantum Optics and Quantum Information. Having the structure of a two-level system, it has been widely used in experiments to test Bell inequalities \cite{Mermin93,Brunner14}, it has inspired the creation of the paradigmatic BB84 quantum cryptography protocol \cite {Gisin02}, and it was used to demonstrate quantum teleportation \cite{bouwmeester97}, superdense coding \cite{mattle96}, one-way quantum computing \cite{walther05}, and several other interesting experiments. 

More recently, we have witnessed the increasing interest in the field of Quantum Thermodynamics, which attempts to combine concepts from Thermodynamics and Classical and Quantum Information theories \cite{Uzdin15,Lostaglio15}. Several experimental platforms have been demonstrated for testing new ideas in this field \cite{bustamante2002,kiang07,sano10,jarion14,Araujo18}, including the photonic polarization \cite{Zanin19}.

The interaction between a single qubit and a thermal bath was interpreted as a Generalized Amplitude Damping (GAD) channel by S. Jevtic \textit{et al.} to investigate quantum thermometers \cite{jevtic15}. The experimental implementation of this model of interaction using photonic polarization was presented by  W. K. Tham \textit{et al.} \cite{Tham2016} and L. Mancino \textit{et al.} \cite{Mancino2017}. This quantum channel is often understood as the interaction between a qubit and an environment with arbitrary temperature. While the GAD also allows control of the interaction time between qubit and environment, we realize here a simplified scheme where the interaction time is set as infinite. We interpret it as a thermal bath with tunable temperature that acts on the qubit. The experimental scheme is simple, reliable and can be used in the study of quantum systems undergoing thermodynamic processes and heat engines.

\section{Photon polarization Thermal States}

We can associate a thermal state, often called Gibbs state, to two-level systems like the
photon polarization. Choosing the vertical ($V$) polarization state to represent the excited
state with energy $\epsilon_2$, and the horizontal ($H$) polarization state to represent the ground state
with energy $\epsilon_1$, the thermal state is written as the density operator:

\begin{eqnarray}
\label{eq1}
\hat{\rho}_\textrm{thermal} &=& P_H |H \rangle\langle H| + P_V |V \rangle\langle V|  \\ \nonumber
&=&  \frac{\e^{-\beta\epsilon_1}}{Z} |\epsilon_1 \rangle\langle \epsilon_1| +  \frac{\e^{-\beta\epsilon_2}}{Z} |\epsilon_2 \rangle\langle \epsilon_2|,
\end{eqnarray}
where $P_H=(1-P_V) \in [0,1]$, $\beta = 1/k_B T$, $k_B$ is the Boltzmann constant, $T$ is the temperature and $Z~=~\Sigma_ie^{-\beta\epsilon_i}$ acts as the partition function of the system. 

The identification between the two-level system thermal state and the polarization mixed state in Eq. (\ref{eq1}) allows us to
obtain the effective inverse temperature $\beta$ as a function of the populations $P_H$ and $P_V$:

\begin{equation}
\label{eq2}
\beta = \varepsilon^{-1} \ln\,(P_H/P_V),
\end{equation}
where $\varepsilon=(\epsilon_2-\epsilon_1)$ is the qubit energy gap. 

\section{Generalized Amplitude Damping Channel}

The action of an environment at some fixed temperature acting on a qubit system is described by the action of the completely positive and trace preserving map $\mathcal{E}$: 
\begin{equation}
\hat{\rho}_{\mbox{out}} = \mathcal{E}\left(\hat{\rho}_{\mbox{in}}\right) = \sum_{k=0}^{3}\hat{\Gamma}_{k}\,\hat{\rho}_{\mbox{in}}\,\hat{\Gamma}_{k}^{\dagger},
\end{equation}
where $\hat{\rho}_{\mbox{in}}$ and $\hat{\rho}_{\mbox{out}}$ are the input and output states of the system and the Kraus operators $\hat{\Gamma}_{k}$ are given by
\begin{eqnarray}
&\hat{\Gamma}_{0}& = \sqrt{1-\xi}\begin{bmatrix}
1 & 0 \\
0 & \sqrt{1-p}
\end{bmatrix}, \hspace{0.3cm} 
\hat{\Gamma}_{2} = \sqrt{\xi}\begin{bmatrix}
0 & 0 \\
\sqrt{p} & 0
\end{bmatrix},  \nonumber \\
&\hat{\Gamma}_{1}& = \sqrt{1-\xi}\begin{bmatrix}
0 & \sqrt{p}\\
0 & 0
\end{bmatrix},\hspace{0.3cm} 
\hat{\Gamma}_{3} = \sqrt{\xi}\begin{bmatrix}
\sqrt{1-p} & 0 \\
0 & 1
\end{bmatrix}.
\end{eqnarray}
where $\xi \in [0,0.5]$ and $p = (1 - e^{-\lambda t}) \in [0,1]$ is the decay probability associated with the excited state, while $\lambda$ stands for the damping constant. This association is useful when the two-level system is related to some physical quantity that is subject to spontaneous decay. However, this is not the case for the photon polarization. 

If we consider an initial state in the general form
\begin{equation}
\hat{\rho}_{\mbox{in}} =  \begin{bmatrix}
\rho_{00} & \rho_{01} \\
\rho_{10} & \rho_{11}
\end{bmatrix},
\label{app:eq:instate}
\end{equation}
the output state will be
\begin{equation}
\hat{\rho}_{\mbox{out}} = \begin{bmatrix}
1-p\xi - \rho_{11}(1-p)  & \rho_{01}\sqrt{1-p} \\
\rho_{10}\sqrt{1-p} & \rho_{11}(1-p)+p\xi
\end{bmatrix},
\label{app:eq:outstate}
\end{equation}
where we have employed the normalization condition $\rho_{00} + \rho_{11} = 1$. Note that in the infinite time limit $t\rightarrow\infty$, i.e., $p\rightarrow 1$, the output state of the channel (GAD$_{t\rightarrow \infty}$) is given by:
\begin{equation}
\hat{\rho}_{\mbox{out}} = \begin{bmatrix}
1-\xi & 0 \\
0 & \xi 
\end{bmatrix},
\label{infinite}
\end{equation}
and $\xi = (1 + e^{\beta \varepsilon})^{-1}$ is now the thermal population of the excited state, with $\beta$ being the inverse temperature defined in Eq. (1). In what follows, we describe the experimental implementation of this channel.

\section{Implementing the photonic GAD$_{t\rightarrow\infty}$ channel}

\begin{figure*}
    \centering
    \includegraphics[width=15cm]{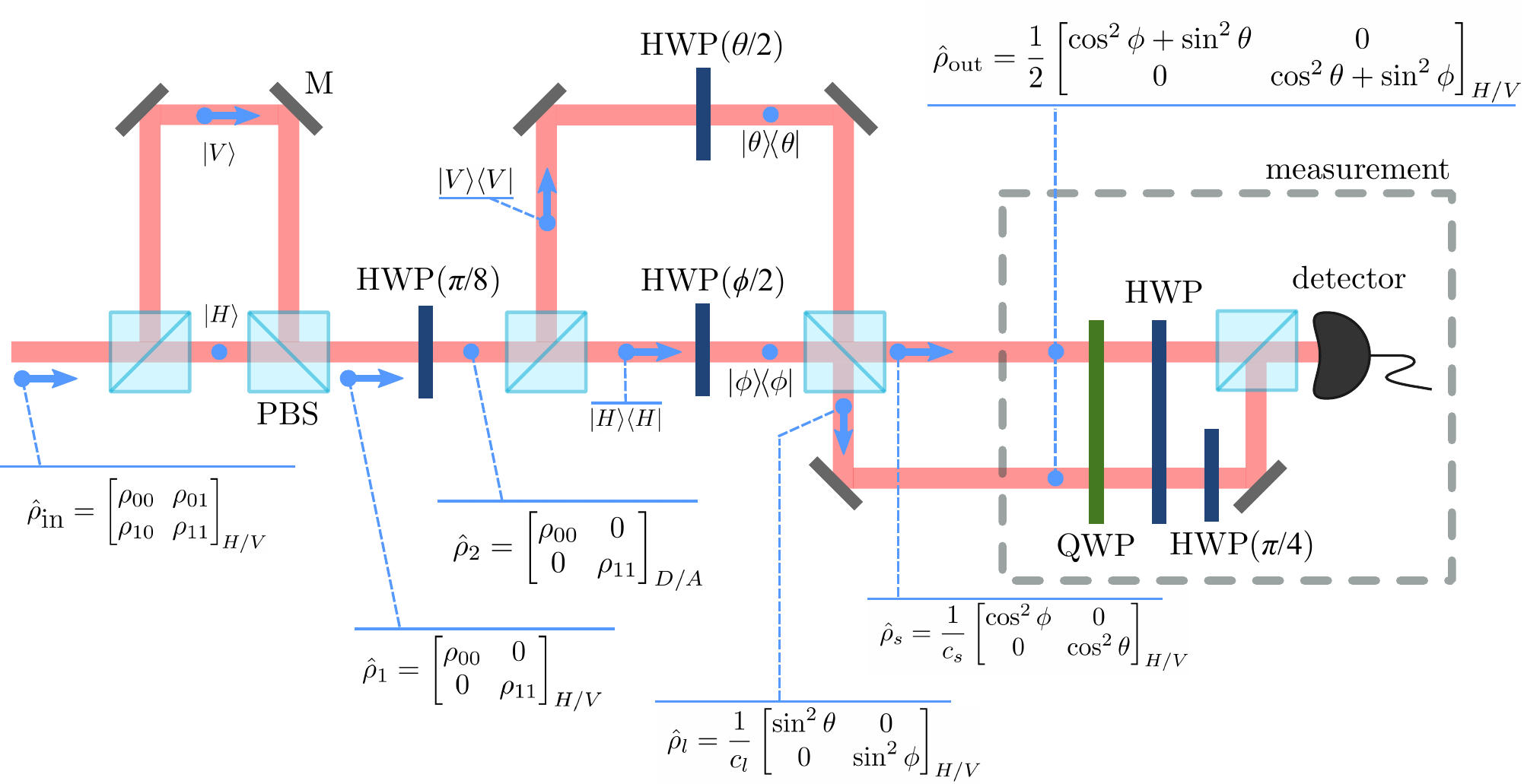} 
     \caption{Scheme for implementing 
     the GAD$_{t\rightarrow \infty}$. See main text for details.}
       \label{fig1}
\end{figure*}

A possible experimental scheme for photonic qubits is displayed in Fig. \ref{fig1}. We will show that it implements the GAD$_{t \rightarrow \infty}$.
Let us consider that the input state is incident from the left and is described by the general density matrix of Eq. (\ref{app:eq:instate}) in the $H$-$V$ basis. The state travels through the first unbalanced Mach-Zehnder interferometer. The path difference is set to be larger than the coherence length of the input state, so that the coherences of the state are destroyed. Therefore, the state evolves to:

\begin{equation}
\hat{\rho}_{1} =  \begin{bmatrix}
\rho_{00} & 0 \\
0 & \rho_{11}
\end{bmatrix}_{H/V}.
\label{roum}
\end{equation}

In the next step it is transmitted through a half-wave plate (HWP) that rotates the polarization states $|H\rangle \rightarrow |D\rangle$ (diagonal) and $|V\rangle \rightarrow |A\rangle$ (anti-diagonal), so that the state is transformed to:

\begin{equation}
\hat{\rho}_{2} =  \begin{bmatrix}
\rho_{00} & 0 \\
0 & \rho_{11}
\end{bmatrix}_{D/A},
\label{rodois}
\end{equation}
where the basis was changed to diagonal/anti-diagonal (D/A).

Next, the state enters the second unbalanced interferometer through a polarizing beam splitter (PBS) that splits the polarization components. In each arm of the interferometer there is a half-wave plate that makes the operation $|V\rangle \rightarrow |\theta \rangle$ and $|H\rangle \rightarrow |\phi \rangle$. The polarization components are recombined in the second PBS. Since the interferometer is unbalanced, the recombination is incoherent and the states at the outputs of the PBS are mixed states given by:

\begin{equation}
\hat{\rho}_{s} =  \frac{1}{c_{s}}\begin{bmatrix}
\cos^2\phi & 0 \\
0 & \cos^2\theta
\end{bmatrix}_{H/V};
\label{roumzero}
\end{equation}
\begin{equation}
\hat{\rho}_{l} =  \frac{1}{c_{l}} \begin{bmatrix}
\sin^2\theta & 0 \\
0 & \sin^2\phi
\end{bmatrix}_{H/V}.
\label{rozeroum}
\end{equation}
where $c_{s}$ and $c_{l}$ are normalization factors:
\begin{eqnarray}
\label{eq801}
&c_{s} = \cos^2\phi + \cos^2\theta; \\ \nonumber
&c_{l} = \sin^2\phi + \sin^2\theta.
\end{eqnarray}

After the interferometer, the trace over the spatial degrees of freedom is made by detection of both outputs with the same detector. In practice, we use the measurement scheme described in Ref. \cite{Salles08}. The final state is given by
\begin{equation}
\hat{\rho}_\textrm{out} = \frac{1}{2} \begin{bmatrix}
\cos^2\phi +\sin^2\theta & 0 \\
0 & \cos^2\theta + \sin^2\phi
\end{bmatrix}_{H/V},
\label{rototal}
\end{equation}
\noindent and the inverse temperature (assuming $\varepsilon=1$) is:
\begin{equation}
\label{eq9a}
\beta_\textrm{out}=\ln \left(\frac{\cos^2\phi +\sin^2\theta}{\cos^2\theta + \sin^2\phi} \right).
\end{equation}
The identification with the action of the GAD$_{t\rightarrow \infty}$ channel  is made by comparing the density matrices in Eqs. (\ref{infinite}) and (\ref{rototal}):
\begin{eqnarray}
1 - \xi = (\cos^2\phi + \sin^2\theta)/2 \, ; \\ \nonumber
\xi = (\cos^2\theta+\sin^2\phi)/2.
\label{GAD}
\end{eqnarray}

Here, we implement a variant of this scheme employing photons produced by a polarization-entangled photon source. In our scheme, the input photonic qubits are prepared in thermal states with easily controllable populations, without the need of the first unbalanced interferometer of Fig. 1.

\section{Experiment}

\begin{figure*}
    \centering
    \includegraphics[width=15cm]{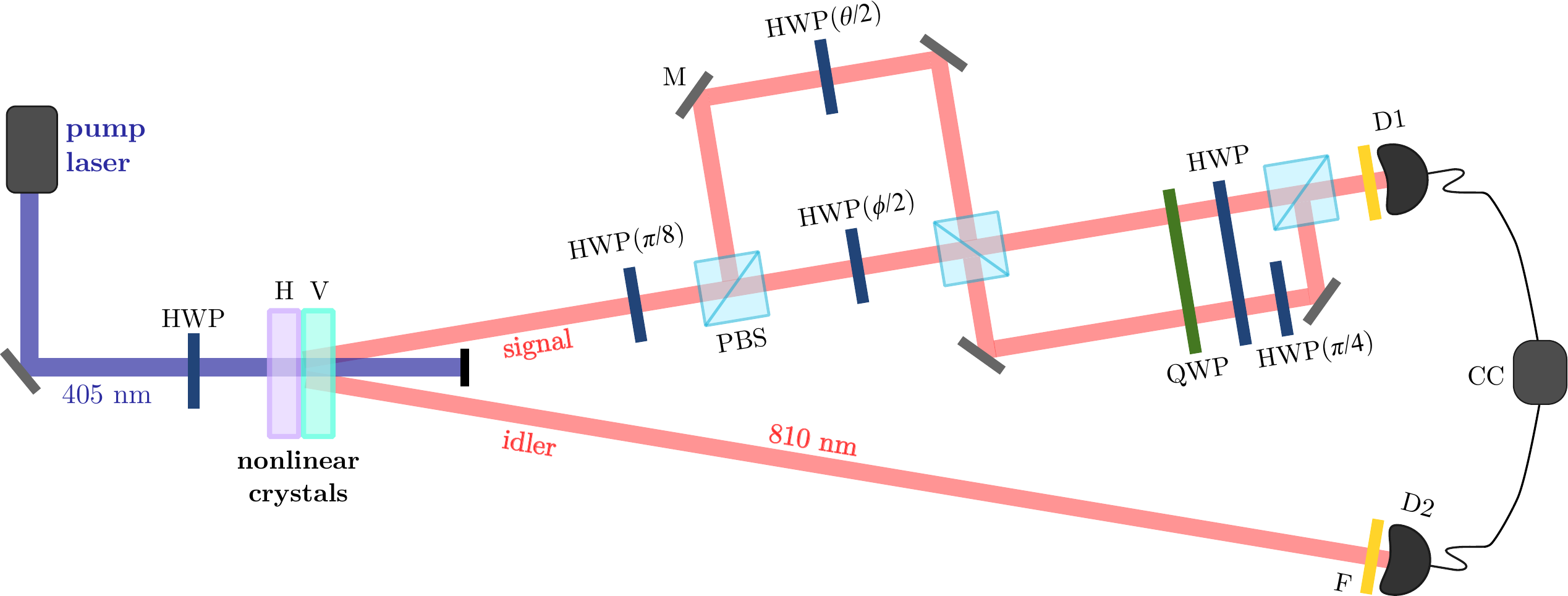} 
     \caption{Experimental setup. The two-crystal spontaneous parametric down-conversion source is pumped by a blue laser and signal and idler photons are emitted. The idler photon is directed to a single photon counting module and the signal photon is directed to two nested unbalanced Mach-Zehnder interferometers.}
       \label{fig2}
\end{figure*}

The experimental setup is shown in Fig.~\ref{fig2}. A diode laser oscillating at 405 nm is used to pump a two-crystal-sandwich spontaneous parametric down-conversion (SPDC) source. This configuration is frequently used as a source of polarization-entangled photon pairs \cite{kwiat99}. The idler photon is directed to a single-photon counting module (SPCM) and detected without measuring its polarization, therefore tracing out this degree of freedom. In this case, the detection of an idler photon heralds a signal photon remotely prepared in a mixed state for the polarization. As we have seen, a mixed polarization state can be interpreted as a thermal state for which the temperature depends only on the populations of linear vertical (V) and horizontal (H) polarization states. 

The SPDC source consists of two Beta-Barium-Borate (BBO) nonlinear crystals. Signal and idler photons are collected along directions such that they have the same wavelength of 810 nm and interference filters centered at this wavelength and having 10-nm bandwidth are placed in front of the SPCMs.

In the path of the idler beam there are two nested, unbalanced Mach-Zehnder interferometers working in the non-interfering regime, because the path length differences are bigger than the coherence length of the signal and idler photons. The optical devices indicated in Fig.~\ref{fig2} are half-wave plates (HWP), mirrors (M), polarizing beam splitters (PBS), interference filters (F), SPCM detectors (D1 and D2) and coincidence counting electronics (CC), while (H) and (V) indicate the pump beam polarization direction that interacts with each crystal.

The overall scheme works as follows. The blue laser pumps the crystals, producing entangled photon pairs. The idler photons are detected and their polarization degrees of freedom are traced out. This prepares the signal photons remotely in a mixed state. By adjusting the pump half-wave plate before the crystals, it is possible to control the populations of the idler photon density matrix and therefore the temperature of the corresponding thermal state. In our experiment we will prepare all input states for the quantum channel in an initial thermal state and in this case the first unbalanced interferometer in Fig.~\ref{fig1} is not necessary. The heralded signal photons prepared in thermal states are incident in the sequence of two nested, unbalanced Mach-Zehnder interferometers that implement the GAD$_{t\rightarrow\infty}$ and polarization analysis.

\section{Results}

\begin{figure}
    \centering
    \hspace*{-0.5cm}
    \includegraphics[width=10cm]{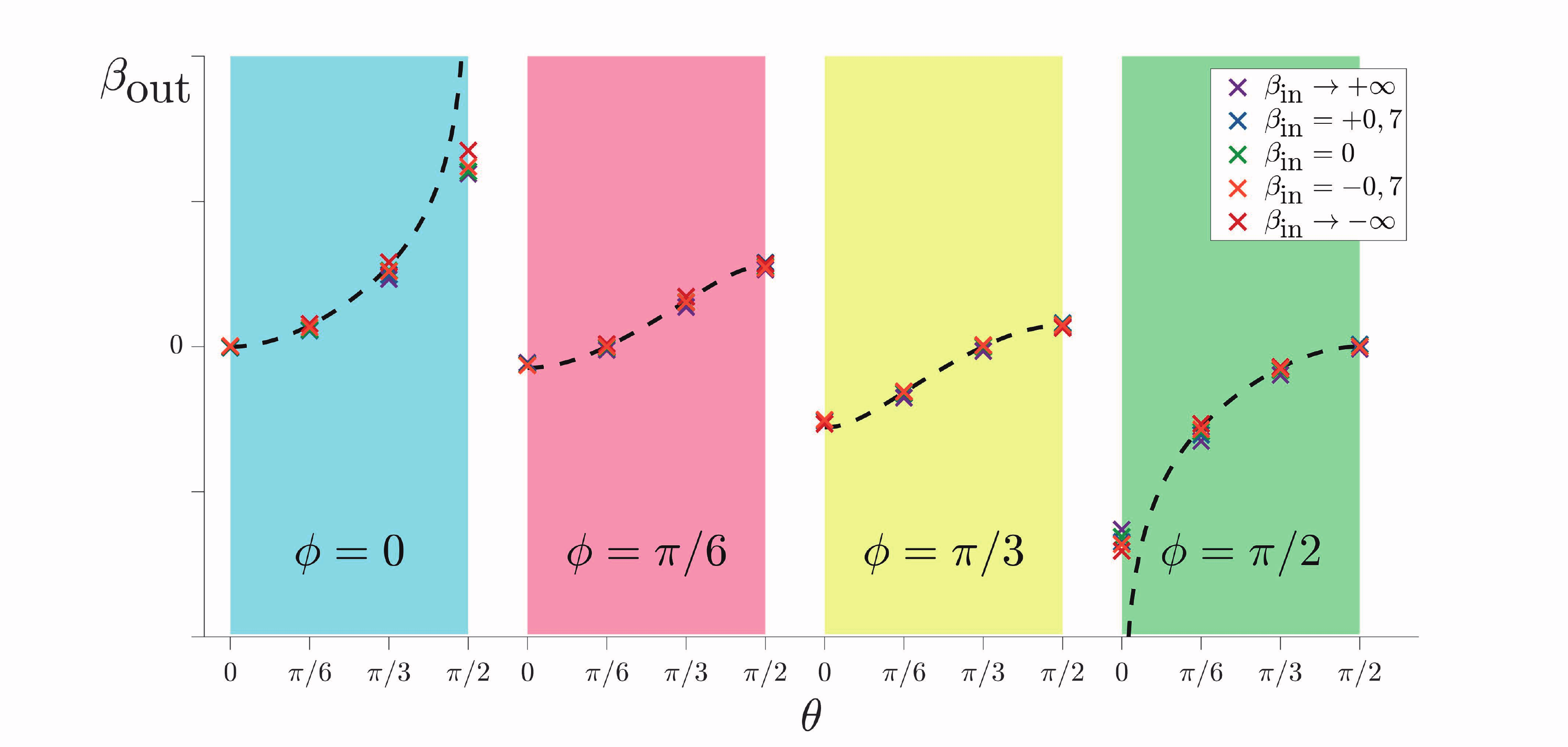} 
     \caption{Experimental results. The effective inverse  temperature $\beta_\textrm{out}$ of the output states after interaction with the heat bath is plotted as function of $\theta$ for four values of $\phi$, and five input states.}
       \label{fig3}
\end{figure}

The experimental results are displayed in Fig. \ref{fig3}. The effective inverse temperature $\beta_\textrm{out}$ is plotted as a function of $\theta$ for four settings of $\phi$. Five different input states were prepared in thermal states with different effective inverse temperatures, including the pure states $|H \rangle$ ($\beta = \infty$) and $|V \rangle$ ($\beta = -\infty$), and the measured output inverse temperatures are shown. The results indicate that the output temperatures are independent of the input state, as expected for the interaction with a heat bath during a time tending to infinite. We note that there are negative temperatures, because in this system it is very easy to have population inversion, in the sense that we can easily prepare a state with larger population in the excited state $|V \rangle$ than the ground state $|H \rangle$.

Even though two-level systems have a simple description and the photon polarization can be controlled and measured with relative simplicity, the experimental configuration that implements a heat bath with controlled temperature is not obvious. Our results show that the GAD is a good option for accomplishing this task. Moreover, in the cases where the dynamics of the coherences do not play an important role, we show that the GAD$_{t\rightarrow\infty}$ works very well. For instance, for realizing heat engines using the photon polarization as the working substance, one can use the GAD$_{t\rightarrow\infty}$. For more general applications, the GAD with variable interaction time is necessary and the experimental implementation requires one additional qubit.
\\


\section{Conclusion}

In conclusion, we present an experimental realization of a photonic qubit quantum channel that acts like a heat bath with controlled temperature and infinite interaction time. We perform measurements using heralded single photons that illustrate the operation of the scheme in the context of this interpretation. We test five different input thermal states, including the pure states $|H \rangle$ and $|V \rangle$ and we find that at the output the temperature of the system is only determined by the parameters of the heat bath. We believe that this setup can be a useful tool in the study of thermodynamic processes in the limit of small scale quantum systems.
\\

\begin{acknowledgements}
The authors would like to thank the Brazilian Agencies CNPq, FAPESC, FAPEG, and the Brazilian National Institute of Science and Technology of Quantum Information (INCT/IQ). This study was financed in part by the Coordenação de Aperfeiçoamento de Pessoal de Nível Superior - Brasil (CAPES) - Finance Code 001. LCC would like to also acknowledge support from Spanish MCIU/AEI/FEDER (PGC2018-095113-B-I00), Basque Government IT986-16, the projects QMiCS (820505) and OpenSuperQ (820363) of the EU Flagship on Quantum Technologies and the EU FET Open Grant Quromorphic and the U.S. Department of Energy, Office of Science, Office of Advanced Scientific Computing Research (ASCR) quantum algorithm teams program, under field work proposal number ERKJ333.
\end{acknowledgements}

\bibliographystyle{apsrev}
\bibliography{bibliographybib}

\end{document}